\documentclass[11pt]{article}

\usepackage[iso-8859-7]{inputenc}
\usepackage{epsfig}
\usepackage{graphicx}

\usepackage{amsfonts}
\usepackage{amssymb}
\usepackage{amsthm}
\usepackage{amsmath}
\usepackage{multirow}

\usepackage{mathtools}

\newcommand{\newc}{\newcommand}
\newc{\ra}{\rightarrow}
\newc{\lra}{\leftrightarrow}
\newc{\ov}{\overline}
\newc{\pa}{\partial}

\newc{\be}{\begin{equation}}
\newc{\ee}{\end{equation}}

\newc{\ba}{\begin{eqnarray}}
\newc{\ea}{\end{eqnarray}}

\newc{\D}{\Delta}

\newc{\la}{\lambda}
\newc{\vt}\vartheta
\newc{\nn}{\nonumber}

\begin{document}
\thispagestyle{empty}

\vskip 2truecm
\vspace*{3cm}
\begin{center}
{\large{\bf
On Topological Modifications of Newton's Law}}

\vspace*{1cm}
{\bf E.G. Floratos~$^{a,b}$, G.K. Leontaris~$^{c}$}\\[3.5mm]
  \normalsize\sl $^a$ Physics Department, University of Athens,
 Zografou 157 84 Athens, Greece \\[2.5mm]
  \normalsize\sl $^b$ Institute of Nuclear Physics, NCSR Demokritos, 15310, Athens, Greece
 \\[2.5mm]
  \normalsize\sl $^c$Theoretical Physics Division, Ioannina University, GR-45110 Ioannina, Greece\\[2.5mm]
\end{center}

\vspace*{1cm}
\begin{center}
{\bf Abstract}
\end{center}

\noindent

Recent  cosmological data  for very large distances challenge
the validity of the standard cosmological model.
Motivated by the observed spatial flatness  the accelerating expansion and
the various anisotropies with preferred axes in the universe we examine
the consequences of the simple hypothesis that the three-dimensional space
has a global ${\cal R}^2\times {\cal S}^1$ topology. We take the radius
of the compactification to be the observed cosmological scale beyond which the
accelerated expansion starts.
 We derive the induced  corrections to the Newton's gravitational potential
and we find that for distances smaller  than the $ {\cal S}^1$ radius
the leading $1/r$-term is corrected by convergent power series of multipole form
 in the polar angle making explicit the induced anisotropy by the compactified
 third dimension. On the other hand,  for distances larger than the compactification scale
  the asymptotic behavior of the potential exhibits a logarithmic dependence with exponentially
  small corrections. The change of Newton's force from $1/r^2$  to $1/r$ behavior
   implies a weakening of the deceleration for the expanding universe.
   Such topologies can also be created locally by standard Newtonian axially symmetric mass distributions
   with periodicity along the symmetry axis. In such cases we can use our results to
   obtain measurable modifications of Newtonian orbits for small distances and flat rotation spectra, for
   large distances at the galactic level.

\vfill
\newpage
\section{Introduction}

Nowadays  there is accumulating evidence that the standard cosmological
model at very large distances might not be fully consistent with the
astronomical data~\cite{Tegmark:2003ve}-\cite{Clifton:2011jh}.
Indeed,  observations  at very large scales  point to  the  existence
 of a possible anisotropy in the global structure of the universe. Similarly,
cosmological data for quasars  indicate that the linear polarization angles of the
emitted photons are correlated to a preferred axis at huge distances of the
order of 1.5 Gpcs~\cite{Hutsemekers:2005iz}. In fact, there is  evidence for
a preferred cosmological axis when comparing accelerations and large peculiar
galaxy velocities at different directions.
Further, from CMB anisotropies observed at the lowest multipoles in
the spherical harmonic decomposition of temperature, there is a hint
of the existence of preferred axis\cite{Tegmark:2003ve}.  The explanation of
rotational curves of galaxies implies the possible existence of dark matter  but
eventually there are other explanations~\cite{Milgrom:1983zz,Sanders:2002pf}.
The standard tests on the validity of Newton's law and general relativity gravitational
corrections are confined inside the solar system, binary pulsars, and gravitational lensing.

The hypothesis that the Universe has a non-trivial topology and it is possibly
 multi-connected has been considered since the beginning of the 20th
 century~\cite{old}\footnote{For recent surveys see for example \cite{LachiezeRey:1995kj}.}.
 Further, similar ideas were developed afterwards~\cite{Zeldovich:1984vk} where a non-trivial
  topology  was also assumed in considerations with regard to the quantum creation of the
  Universe and the origin of inflation~\cite{Vilenkin:1982de,Easson:2010av}.

In the present note, we consider the possibility that  matter does not only influence
the surrounding space-time geometry but also the topology in various scales.
This idea comes from the well known fact, that in order to describe local
physics we need to select a complete set of smooth functions on the whole space
in order to form delta functions, Greens functions etc, but then the topology of
the space is crucial in determining this complete set.
One possible and  simple approach is to consider violations of Newton's law emerging
from local changes of the topology of the Euclidean space ${\cal R}^3$ to
${\cal R}^2\times {\cal S}^1$. In a subsequent work we will develop further this
idea in the framework of Einstein's Theory of General Relativity.
At present,  we will confine our analysis on the Newtonian limit of the above idea i.e.
 the global topology change of ${\cal R}^3$
Euclidean space to ${\cal R}^2\times  {\cal S}^1$.
 We will obtain then, the exact solution for
the Newton's potential interpolating between short distances where the usual
$\frac{1}{r}$-behavior appears, to the large distance behavior $\log r$
for distances much larger the radius $R$ of $ {\cal S}^1$.   We will see that
 we obtain axial symmetric solutions with multipole terms and
 indications of correct qualitative behavior for the rotational spectra of galaxies.
 Other tests like gravitational lensing and spectra of CMB~\cite{Urban:2010wa}
 require further studies.

   We find also that our result is   equivalent to that obtained  from an infinite  discrete and periodic
  mass distribution along the decompactified dimension.
  An analysis of these implications on the Keplerian orbits   of Stars at distances where the first non-leading corrections are important, is also presented.  A detailed comparison with the astrophysical data and the observed uncertainties in beyond the scope of this letter and will be the  subject of a future publication. In the
  present note,  in section 2 we derive and discuss in detail the properties of the ${\cal R}^2\times  {\cal S}^1$
  potential and in section 3 we analyse its implications on the trajectories of massive objects. The results are discussed in section 4.

\section{Gravitational Potential in ${\cal R}^2\times  {\cal S}^1$}

Models for modifications of Newton's Gravity have already been proposed for  large as well  as
for minuscule distances, the latter being only relevant in the presence of extra compact dimensions~\cite{Floratos:1999bv}. A survey of  recent  possibilities of modifications
of the Newtonian Gravity can be found in~\cite{Sanders:2002pf,Clifton:2011jh}\footnote{For
 other types of modified gravity scenarios related to dark matter, the accelerating expansion of the
 universe  etc see for example~\cite{Nojiri:2006ri,Copeland:2006wr,Woodard:2006nt}.}.
 Here, in this work we perform a mathematical analysis which will prove useful to examine whether
 in the absence of Dark Matter, such modifications could also be interpreted as a possible effect
 of the ordinary matter in the topology of the space.

We assume a three dimensional space with the topology ${\cal R}^2\times  {\cal S}^1$.
 We parametrise the  compact dimension of `radius' $R$ as $z=R\,q$ with $q=[0,2\pi)$.
 The  Poisson  equation  for a point unit mass  in this case is written
\ba
\nabla^2_{2+1}\Phi&=&\delta^{(2)}(\vec \tau-\vec \tau_0)\frac{1}{R}\delta(q-q_0)
\ea
with $\vec \tau=(x,y)\in \mathbb{R}^2$. Expanding  in Fourier modes, the solution
can be expressed as follows
\ba
\Phi&=&-\frac{1}{(2\pi)^3R}\sum_{m=-\infty}^{\infty}\int\,d^2k\,e^{i\vec
k\cdot\vec \tau+\imath
m\,q}\int_0^{\infty}ds\,e^{-s(k^2+\frac{m^2}{R^2})}\label{infsum}
\ea
Performing the $\vec k$  Gaussian integration
 the potential takes the form
\ba
\Phi
&=&-\frac{1}{8\pi^2R}\sum_{m=-\infty}^{\infty}e^{\imath
m\,q}\int_0^{\infty}\,e^{-\frac{\tau^2}{4
s}-s\frac{m^2}{R^2}}\,\frac{ds}{s} \label{infsum3}
\ea
To proceed  with our computation, we split the infinite sum (\ref{infsum3}) into two parts
\ba
\Phi&=&-\frac{1}{8\pi^2R}\left[\int_0^{\infty}\,e^{-\frac{\tau^2}{4
s}}\,\frac{ds}{s}+2\sum_{m=1}^{\infty}\cos\,m\,q\,\int_0^{\infty}\,e^{-\frac{\tau^2}{4
s}-s\frac{m^2}{R^2}}\,\frac{ds}{s}\right]\;=\;\Phi_0+\delta \Phi\label{2parts}
\ea
 The first integral diverges, however, we observe that it is the result  of the Fourier
 integration of the zero-mode
\ba
\Phi_0&=&-\frac{1}{8\pi^2R}\int\,d^2k\, e^{\imath\vec k\cdot\vec
\tau}\int_0^{\infty}ds\,e^{-sk^2}\;=\;-\frac{1}{8\pi^2R}\int\,d^2k\,
\frac{e^{\imath\vec k\cdot\vec \tau}}{k^2}
\ea
This obeys the two-dimensional Poisson equation
\ba
\nabla^2\Phi_0&=&\frac{1}{2\,R}\;\delta^{(2)}(\vec \tau)
\ea
with the known solution
\ba
\Phi_0&=&\frac{1}{4\pi^2}\,\frac 1R\,\ln\frac{\tau}{\tau_0}
\ea
The result of the second  integration in (\ref{2parts}) is denoted with $\delta \Phi$
 and it equals twice the modified Bessel function $2\,K_0(m\tau/R)$.
Hence, adding the two contributions the potential takes the form
\ba
\Phi&=&\frac{1}{4\pi^2R}\,\left(\ln\frac{\tau}{\tau_0}-2\sum_{m=1}^{\infty}\cos\,(m\,q)
\,K_0(m\tau/R)\right)
\label{AllV}
\ea
We can obtain a closed form of the second term using a known formula from \cite{Gradshteyn}.
Plugging in  the appropriate variables, we get
\begin{equation}
\begin{split}
\sum_{m=1}^{\infty}K_0(m \frac \tau R)\,\cos\,m\,q&=\frac 12
(\gamma+\ln\frac{\tau}{4\pi R})+\frac{\pi
R}{2\sqrt{\tau^2+(R\,q)^2}}\nonumber
\\
&+\frac{\pi
}{2}\sum_{l=1}^{\infty}\left[\frac{R}{\sqrt{\tau^2+R^2(2\pi
l-\,q)^2}}-\frac{1}{2\pi l}+\frac{R}{\sqrt{\tau^2+R^2(2\pi
l+\,q)^2}}-\frac{1}{2\pi l}\right]
\end{split}
\end{equation}
 Using this expansion we notice that the logarithms   in (\ref{AllV})
cancel each other.  We may further choose the constant $\tau_0$ to be
$\tau_0=4\pi\,R\,e^{-\gamma}\;\approx\; 2.246\,\pi\,R$
so the final form of the potential valid for all distances  takes the form
\ba
\Phi&=&-\frac{1}{4\pi}\frac{1}{\sqrt{\tau^2+R^2\,q^2}}\nonumber
\\
&-&\frac{1}{4\pi}\sum_{l=1}^{\infty}\left[\frac{1}{\sqrt{\tau^2+R^2(2\pi
l-\,q)^2}}-\frac{1}{2\pi l R}+\frac{1}{\sqrt{\tau^2+R^2(2\pi
l+\,q)^2}}-\frac{1}{2\pi l R}\right]
\label{pot2PP1}
\ea
From this latter expression, we observe   that the three dimensional Newton's
 potential is recovered in the limit $R\ra \infty$,  $q\to 0$ with $z=R\,q $ finite.
 We further observe that  form of the potential in (\ref{pot2PP1}) can also be represented
as the Newtonian potential of an infinite discrete
mass distribution of  period $2\pi R$ along the decompactified  dimension.

\subsection{The potential at short distances}

The above form of the potential encodes the axial symmetry of the topology of space and
therefore it  is interesting to investigate  particular limiting cases
 with respect  to the coordinate axes. However it is not easy to  elaborate it further
 in its present form. In order to do this, we need to introduce  a
convenient parametrization.
To this end, we define
\begin{equation}
\begin{split}
a_l&=\;\frac{r}{2\pi\,l\,R}
\end{split}
\end{equation}
and we note that $z =Rq= r\cos\theta$. Now we can expand the square roots  into
Legendre polynomials for small $r/(2\pi R)$
\ba
\frac{1}{\sqrt{\tau^2+R^2(2\pi l\pm \,q)^2}}&=& \frac{1}{2\pi
l R}\frac{1}{ \sqrt{1+a_l^2\pm 2a_l\,\cos\theta}}\nonumber
\\
&=&\frac{1}{2\pi lR}\,\sum_{k=0}^{\infty}P_k(\pm\cos\theta)\,a_l^k
\ea
Substituting into  the series form of $\Phi$ while using the
property  $P_k(-\cos\theta)=(-)^k P_k(\cos\theta)$ and changing
the order of the two summations we obtain
\ba
\Phi&=&-\frac{k}{4\pi}\frac{1}{r}
-\frac{k}{2\pi}\sum_{l=1}^{\infty}\frac{1}{2\pi
lR}\left[\sum_{n=0}^{\infty} \frac{r^{2n}}{(2\pi lR)^{2n}}P_{2n}(\cos\theta)-1\right]\nn
\\
&=&-\frac{k}{4\pi}\frac{1}{r}\left(
1+2\sum_{n=1}^{\infty}\zeta(2n+1)\,\left(\frac{r}{2\pi
R}\right)^{2n+1} P_{2n}(\cos\theta)\right)\label{PotC}
\ea
We remark first that the above expression for the potential holds
for small distances $r/(2\pi R)<1$  and moreover care should be taken in the
interpretation of the spherical coordinate variables $r,\theta$
in order to implement the topology of the space ${\cal R}^2\times  {\cal S}^1$.

In the following we are going to examine the behavior of the potential
in various directions of $ {\cal R}^2\times {\cal S}^1$ space.
We study   two  cases for the potential, the first
along the $z$-axis ( $\theta=0$) and the second on the $(x,y)$ plane
($\theta=\frac{\pi}2$) .

In the first  case, $(\theta=0)$, we have $P_n(1)=1$ and the infinite sum in (\ref{PotC})
can be expressed in terms of the Euler constant $\gamma$ and PolyGamma functions
so that the potential along the $z$-axis is given in closed form
\ba
V(r)&=&-\frac{k}{r}+\gamma+\frac{k}{\pi R}\left( \psi^{(0)}
\left(1-\frac{r}{s}\right)+\psi^{(0)}\left(1+\frac{r}{s}\right)\right),\;s=2\pi R
\ea
 For the second case $\theta =\frac{\pi}2$ and
\[P_{2n}(0)=\frac{(-1)^n  (2 n)! }{2^{2 n}(n!)^2}\]
   The potential in this case takes the form
\ba
V(r)&=&-\frac{k}{ r}\left[1+{2}\,\sum _{n=1}^{\infty } \frac{(-1)^n (2 n)!
\zeta (2 n+1)}{2^{2 n}(n!)^2}\left(\frac{r}{ 2\pi R}\right)^{2n+1}\right]
\ea
For large $n$ the coefficients of the powers of the expansion parameter $s=r/(2\pi R)$
goes as $(-1)^n/\sqrt{n}$, so we can approximate the above sum by appropriate combinations
of the polylogarithmic functions $\text{Li}_{\frac{1}{2}}$.
\begin{figure}[!b]
\centering
\includegraphics[scale=0.70]{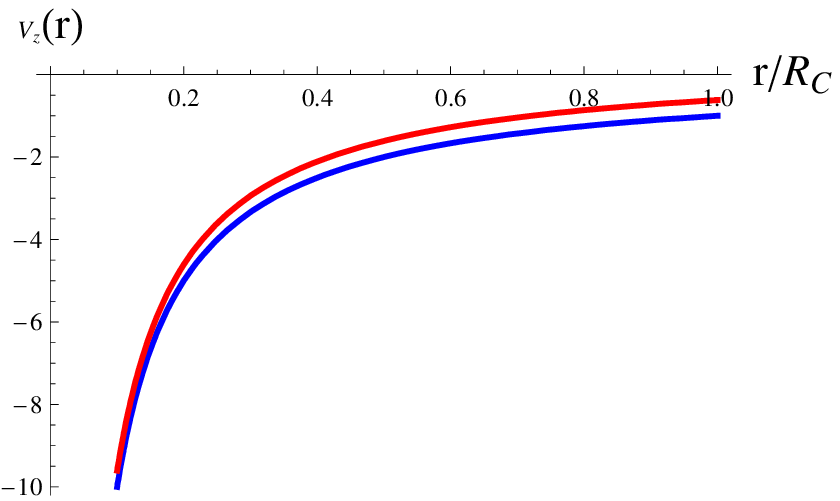}\;\;
\includegraphics[scale=0.55]{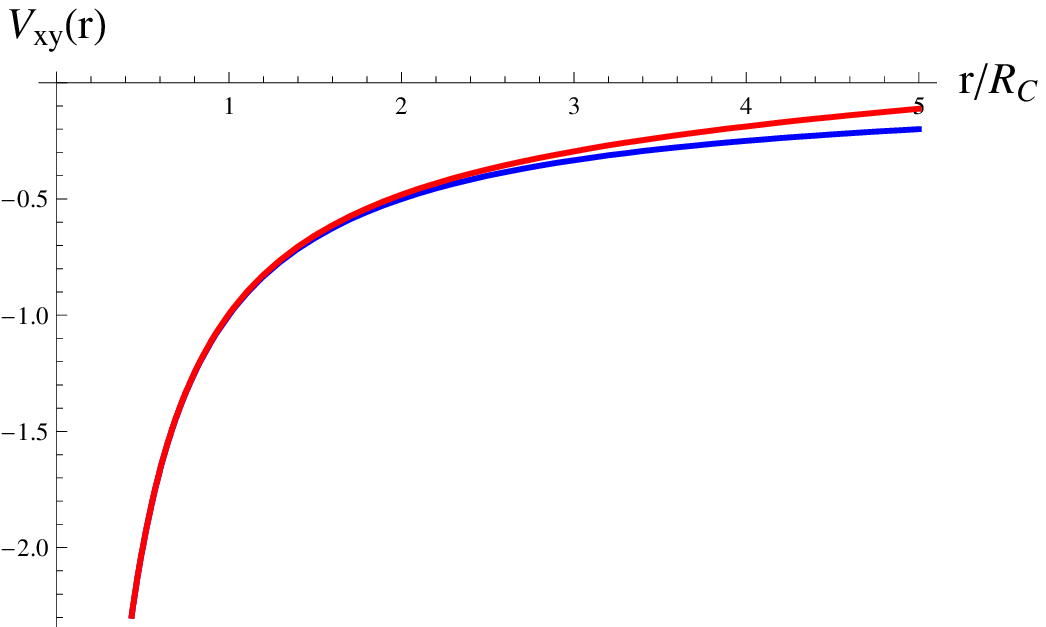}
\caption{Modifications of the Gravitational potential (in arbitrary units) a) along the $z$-axis
b) along the $xy$-plane. Blue (lower) lines
corresponds to ${\cal R}^3$-case, the red ones (upper) to ${\cal R}^2\times  {\cal S}^1$ topology.}
\label{Vz}
\end{figure}
The above considerations describe the behavior of the potential near the decompactification
limit $r\ra 0$.  Figure~\ref{Vz} shows the modification of the gravitational potential for
 ${\cal R}^2\times  {\cal S}^1$ topology in comparison to  ${\cal R}^3$ case.

\subsection{ The potential for large distances}

Because of the compactification radius bound, large distances where $r/(2\pi R)>1$,
can be reached only in the direction $\theta=\frac{\pi}2$ or $q=0$. In this case, expanding
 the Bessel function in (\ref{AllV}) for large values of $t=m\rho$ with $\rho=\tau/R$ we have
\begin{equation}
\begin{split}
K_0(t)&=
\sqrt{\frac{\pi}{2t}}e^{-t}\sum_{n=0}^{\infty}
\frac{[(2n-1)!!]^2}{n!}\frac{1}{(-8t)^n}\\
&=\sqrt{\frac{\pi}{2t}}e^{-t}\left(1-\frac{1}{8t}+\frac{3^2}{2!(8 t)^2}-\frac{(3\cdot 5)^2}{3!}\frac{1}{(8t)^3}+\frac{(3\cdot 5\cdot 7)^2}{4!}\frac{1}{(8t)^4}+\cdots\right)
\end{split}
\end{equation}
and  performing  the sum  in (\ref{AllV}) we  obtain
\begin{equation}
\begin{split}
\sum_{m=1}^{\infty}K_0(m\rho)&=\sqrt{4\pi}\sum_{n=0}^{\infty}\frac{[(2n-1)!!]^2}{n!}
\frac{(-1)^n}{(8\rho)^{n+\frac 12}}{\rm Li}_{n+\frac 12}(e^{-\rho})\\
&=
\sqrt{\frac{\pi }{2\rho}}\left(\text{Li}_{\frac{1}{2}}\left(e^{-\rho}\right)
-\frac{1}{8\rho}\,\text{Li}_{\frac{3}{2}}\left(e^{-\rho}\right)+
\frac{9}{128\rho^2}\,\text{Li}_{\frac{5}{2}}\left(e^{-\rho}\right)+\cdots \right)
\end{split}
\end{equation}
The complete potential is obtained by adding the logarithmic term. This in the
large $\tau/R$ regime  becomes
\begin{equation}
\begin{split}
V_c(\rho)&= \frac{1}{2\pi^2}\frac{1}{R}\left(-C+ \frac 12\log(\rho)-\sqrt{\frac{\pi}{2\rho}}\left(
\text{Li}_{\frac{1}{2}}\left(e^{-\rho}\right)-\frac{1}{8\rho}\text{Li}_{\frac{3}{2}}\left(e^{-\rho}\right)
+\cdots \right)\right)
\end{split}
\end{equation}
where $C$ is the constant
\[C=\ln(2\sqrt{\pi})-\gamma/2\]
We note here that the corrections of the logarithmic term are exponentially small.
\begin{figure}[!b]
\centering
\includegraphics[scale=0.75]{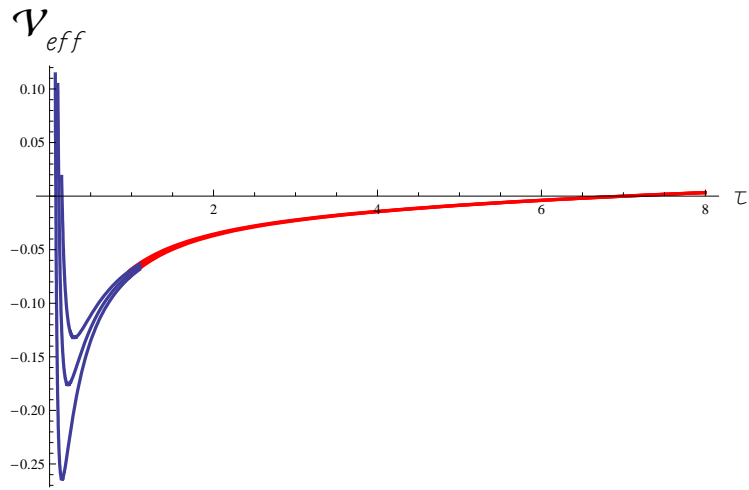}
\caption{Plot of the approximate formulae of ${\cal R}^2\times  {\cal S}^1$
 gravitational potential ${V_{eff}}(\tau/R)$ for small $\tau/R$ ( left blue curves) and large $\tau/R$ (right red curves).}
\label{Vtau}
\end{figure}
In figure~\ref{Vtau}  we plot the effective potential $V_{eff}(\tau)=V_c+\ell^2/(2m\tau^2)$,
taking three values for $\ell/(2 m)=0.06,0.09,0.12$ where $\ell$ is the conserved angular momentum and $m$
the mass. For large values of $\tau/R$ we add  correction terms $\text{Li}_{n+\frac{1}{2}}$ up to $ n=2$.
These correspond to the right piece of  the curves (red color).  At distances around $r\approx 2 \pi R$
the curves cross the  abscissa and for larger $r$-values, the potential becomes positive and increasing.
 Clearly this is due to the logarithmic dependence of the potential. To plot the potential for small $\tau/R$
we use the  formula (\ref{PotC})  taking $n=10$ terms of the sum (in practise only few
of them suffice). These values are represented by the left part of the curves (blue color).
We observe that the two curves match perfectly around the area $\rho\approx 1$, thus
our approximation is valid for small   $\tau/R \lesssim  1$ as well as large  $\tau/R\gtrsim 1$ values.

We observe now that the corrections terms constitute an alternating series of polylogarithmic functions
of the exponential $e^{-\rho}$. Thus increasing the values $\rho >1$, their value decreases and the
dominant contribution comes from the logarithmic term.  So, for large distances the
 rotational spectra of  massive objects are flat
 \[ v\approx {\rm constant}\]
 We notice here that axially symmetric periodic Newtonian mass distributions along the $z$-direction
 can mimic the effect of the compactification ${\cal R}^2\times  {\cal S}^1$. So although presently
 there is no observational basis for such mass distributions, ( for example
 concentration of  galaxies along cosmic strings)
 we can choose the compactification radius $R$ to be of the order of the average intergalactic scale.
This way,  our mechanism  could mimic the effects of the dark matter.

\section{Modification of the orbits}

The modifications of the potential discussed above, are expected to be negligible at small distances.  Indeed, one can notice that the relative coefficient of the first correction term is $\frac{2\zeta(3)}{(2\pi)^3}P_2(0)\approx -0.005$ , thus  for $r\ll R $ its contribution  diminishes. There are considerable deviations from Newton's
mechanics however, when dealing with distances at very large scales.  In this case the relation (\ref{AllV})
 gives the leading logarithmic behavior with  exponentially small corrections.

 At intermediate distances the effect of the corrections will be  important
but we need to know the exact behavior of the potential which, although complicated algebraically,
numerically interpolates nicely between the $1/\tau$ and $\log(\tau)$ behavior.
Here  we examine the effects of the violations of the Newton's law in the $\mathbb{R}^2$-plane
for the orbits of massive objects and for small distances.

For  distances inside the compactification scale,
where the first non-leading correction to the Newtonian law plays a role
it is interesting to consider  the nature of modifications on the
orbits of massive objects. We start with the calculation of the force
taking into account the first
non-leading term and in polar coordinates ($r,\theta,\phi$) we find
\ba
\vec F=\vec F_0+\vec F_c&=&k\left\{-\frac{1}{ r^2}+\frac{3  r
\,\zeta (3) }{8\pi^3 R^3}\left(\cos 2 \theta+\frac 13\right),\,-\frac{3 r\,
    \zeta (3)}{8 \pi ^3 R^3}\sin 2\theta,\,0\right\}\label{ForceAll}
\ea
The form of the correction to Newton's force is similar to the tidal forces on a dipole.
At this level of approximation we observe that the axially symmetric additional forces
are small and repulsive for $\cos(2\theta) >-\frac 13$  and attractive otherwise.
For the study of the effect of this correction to the orbits we restrict
 our attention to the planar motion ($\theta=\frac{\pi}2$), $F_{\theta}=0$
while
\ba
F_r&\approx& -\frac{k}{r^2}-\frac{k\zeta(3)}{4\pi^3R^3}r
\ea
The corresponding  equation of motion  is
\ba
\frac{d^2u}{d\varphi^2}+u&=&\frac{km}{\ell^2}\left(1+\frac{2\zeta(3)}{(2\pi
R)^3}\frac{1}{u^3}\right)\label{DEc}
\ea
where $u= \frac 1r$. This  equation is easily solved  using the conservation
of the angular momentum,
\ba
\varphi-\varphi_0&=&\int_{r_0}^r\frac{d\tau}{\tau^2\sqrt{\frac{2m}{\ell^2}\left({\cal
E}-V(\tau)-\frac{{\ell}^2}{2m\tau^2}\right)}}\label{phi}
\ea
where
\ba
{\ell}&=&mr^2\dot{\varphi}
\ea
 The correction to the Newton's potential is an even-power series of $r$, the most dominant being the
 harmonic term. Thus, the effective potential can be approximated by
\ba
V_{eff}(r)&=&-\frac{k}{r}\left(1-\frac{{\ell}^2}{2mk\,r}-\delta\frac{\zeta(3)}{(2\pi)^3}
\left(\frac{r}{R}\right)^3\right)
\ea
where $\delta=0$ for the classical Newton's law and $\delta =1 $ for the
corrected potential.  The exact computation of the integral would introduce complicated
 elliptic functions, however, since the  correction is small it can be treated perturbatively by
  expanding appropriately the denominator and computing the relevant integrals.
  Instead of this, we will work out in the next section the corresponding equation of motion.

\subsection{The equation of motion}

In the approximation discussed above we may analyse easily the motion using  the standard
approach by deriving  the differential equation for the new variable $u=1/r$  introduced above.
The perturbed differential equation is written as follows
\ba
\frac{d^2u}{d\varphi^2}+u&=&\alpha (1+\epsilon u^{-3})\label{DEper}
\ea
where $\alpha= {km}/{\ell^2}$ and  the  parameter $\epsilon$ can be deduced from
(\ref{DEc})
\[\epsilon=\frac{\zeta(3)}{4\pi^3R^3}\]
 We assume solutions of the form
\ba
u&=&u_0+\epsilon u_1
\ea
where $u_0$ is the classical elliptic solution
\ba
r_0(\varphi)&=&\frac{a(1-e^2)}{1+e\cos(\varphi)}=\frac{1}{u_0(\varphi)}\label{r0c}
\ea
where as usually, $a,e$ stand for the semimajor axis and the
eccentricity~\footnote{These are defined as $a=-\frac{k}{2E}$ and
$e=\sqrt{1-\frac{2E\ell^2}{mk^2}}=\sqrt{1-\frac{\ell^2}{kma}}$.
Note also that the parameter $\alpha$   is  related
to the physical parameters $a,e$ by $\alpha^{-1}=a (1-e^2)$.}.
We seek solutions for the perturbed equation which have the same angular momentum
as the unperturbed one, thus $\delta\ell=0$.
Substituting into (\ref{DEper}) and using the equation (\ref{r0c}) for $u_0(\varphi)$, one finds
\ba
\frac{d^2u_1}{d\varphi^2}+u_1&=& \frac{\alpha}{ u_0^3(\varphi)}\label{DEper1}
\ea
This has the solution
\ba
u_1(\varphi)&=&\frac{I(\varphi)}{\alpha^2}=\,\frac{\cos\varphi
\,I_s(\varphi)-\sin\varphi\,I_c(\varphi)}{\alpha^2}
\ea
 where  $I_s(\varphi)$ and $I_c(\varphi)$ are given by
\begin{equation}
\begin{split}
I_c(\varphi)&=\frac{3 e }{\left(1-e^2\right)^{5/2}}\tan ^{-1}\left(\sqrt{\frac{1-e}{1+e}} \tan
   \frac{\varphi}{2}\right)-\frac{\left(2+e^2+\left(1+2 e^2\right) e \cos \varphi\right)
   \sin \varphi}{2 \left(1-e^2\right)^2 (1+e \cos \varphi)^2}\nonumber
   \\
  I_s(\varphi) &=\frac{1}{2 e}\left(\frac{1}{(1+e)^2}-\frac{1}{(1+e \cos \varphi)^2}\right)
   \end{split}
   \end{equation}
   To get a feeling of  the corrections implied on the trajectories of physical objects
   we form the ratio of the corrected over the unperturbed solution.
    To this end, we expand the ratio in terms of the dimensionless quantity
    $\epsilon/\alpha^3$  and find  that for small eccentricities $e\ll 1$ the essential
    modification of the orbits is captured by the formula
\ba
r(\varphi)\approx r_0(\varphi)\,\left(1- \xi^3\,
\sin^2\frac{\varphi}2\right)\label{solc}
\ea
where $r_0(\varphi)$ is given by (\ref{r0c}) and we have defined
\[ \xi=\left(\frac{\zeta(3)}{2}\right)^{\frac 13}\,\frac{a}{\pi R}\]
 For $R$ of the order of galactic distances and $a$ a typical solar system
distance, this implies a negligible distortion on the orbits.

\begin{figure}[!t]
\centering
\includegraphics[scale=0.9]{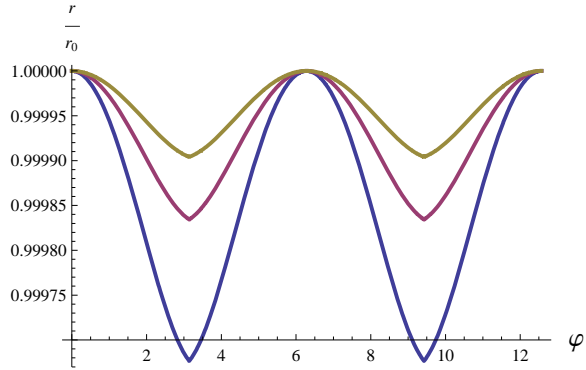}
\caption{ The ratio $\frac{r(\varphi)}{r_0(\varphi)}$  with $r$ being the distorted  orbit
in the presence of the correction term in ${\cal R}^2\times  {\cal S}^1$  topology. The three curves
(upper, middle, lower) correspond to three different ratios $a/R=1/6,1/5,1/4$ where $a$ is the
semimajor axis of the unperturbed orbit and $R$ the  $ {\cal S}^1$-radius.}
\label{Orbit}
\end{figure}

 Next we  study the impact on the trajectories of massive objects with bounded motion,
using three values for the ratio $\lambda=1/6,1/5,1/4$ while assuming  the eccentricity $e=0.068$.
 For an assumed massive object subject to the forces due to a potential generated at the
origin, the corresponding ratios are plotted in figure (\ref{Orbit}).
It is obvious  that for massive objects at the boundaries of galaxies the radius $R$ has to be taken
of the same order to have an observable modification of Newtonian orbits.

We close our analysis with the following remark.
Suppose that in the absence of the ${\cal R}^2\times  {\cal S}^1$-effects the orbit of an object
with  eccentricity  $e$ is determined as usual by $r_0(\varphi)$ given by (\ref{r0c}).
Substituting, the solution (\ref{solc}) can be also written as follows
\[r(\varphi)\approx a (1-e)\frac{1- \xi^3\,
\sin^2\frac{\varphi}2}{1- \frac{2e}{1+e}\,
\sin^2\frac{\varphi}2}\]
For objects satisfying $\xi^3=2e/(1+e)$, or equivalently
\[a=\pi\,R\,\left(\frac{1}{\zeta(3)}\frac{4e}{1+e}\right)^{\frac 13}\]
and to first order approximation on the eccentricity $e$, we obtain
\[r(\phi)=a (1-e) ={\rm constant}\]
Therefore,  certain ``would be''  elliptic orbits in ${\cal R}^3$ space,
in the  case of ${\cal R}^2\times  {\cal S}^1$ topology appear to be approximately circular.

\section{Discussion and conclusions}
  Motivated by the fact that several recent astronomical and
  cosmological observations are challenging the theory of General
Relativity at very large distances, due to accelerated expansion, axial
anisotropies, planarity of CMB multipoles, dark energy etc in this note
we have considered the role of a non-trivial topology of the universe
in the Newtonian approximation without invoking extra dimensions.

Assuming the experimentally determined flatness of the spatial universe,
we test at a qualitative level the proposal of the simplest periodic
multiverse along the third spatial dimension or equivalently the assumption
of a spatial topology ${\cal R}^2\times  {\cal S}^1$ with a radius of the
compactified dimension of the order of the distance where the acceleration
starts, $L=2 \pi R.$

 Within the aforementioned topology of the space,
 we have determined the exact Newtonian potential at any distance $r$.

For small $r/R$ ratios we find the corrections  to the three dimensional
Newton's $1/r$-potential to be of the form of tidal forces $(r/R)^n$
$P_n(\cos\theta)$, with an anisotropy axis  along the third dimension. Another
interesting finding at these distances is the repulsive accelerated motion
along the axial direction in consistency  with the observed large peculiar
velocities of galaxies along the observed axes at some statistical
significance level.

On the other hand, for large distances, where $r/R >1$, we get a smooth
transition of the potential to the two dimensional one, $\log(r)$, with
exponentially small corrections. This has the immediate consequence that
for distances larger than $L$ we obtain  smaller decelerations due to the
appearance of the  two dimensional attractive force $1/r$, instead of  the
$1/r^2$ of three dimensions, thus a relative acceleration.

 Moreover, we expect that for such large distances or angular separations there
must be a flattening of the CMB multipoles due to lower dimensionality of space.

Choosing smaller scales for the compactification radius could be afforded
only in some General Relativity model, where the compactification is local
and not global. Another possibility is to mimic the compactification of the third dimension
with spatial axially symmetric and periodic in the third direction Newtonian
mass distributions. In such a patch -say around a galaxy- we  obtain that the
trajectories of massive stars at the boundaries of the galaxy have flat
rotational spectra and small but measurable perturbations of the
periodic orbits for smaller distances.

We find the above results encouraging and the proposed topological
modification of gravity worth of extending, in a future work, in the
framework of General Relativity.

\vfill

{\bf Acknowledgements} We would like to thank CERN Theory Division for kind hospitality where
part of this work has been done.  The research Project  is co-financed by the European Union - European Social Fund (ESF) \& National Sources, in the framework of the program ``THALIS" of the ``Operational Program Education and Lifelong Learning" of the National Strategic Reference Framework (NSRF) 2007-2013.

\end{document}